\documentclass[pre,twocolumn,showpacs,amsmath,amssymb,floatfix]{revtex4}

\usepackage{graphicx}
\usepackage{ifthen}

\begin{document}

\title{Theory of self-assembled smectic-\textit{A} ``crenellated disks''}

\author{Hao Tu and Robert A. Pelcovits}
\affiliation{Department of Physics, Brown University, Providence RI, 02912, U.S.A}

 \date{\today}

\begin{abstract}
Smectic-\textit{A} monolayers self-assembled in aqueous solutions of chiral \textit{fd} viruses and a polymer depletant have been shown to exhibit a variety of structures including large, flat disks and twisted ribbons. The virus particles twist near the edge of the structure in a direction determined by the chirality of the viruses. When \textit{fd} viruses and their mutants of opposite chirality are mixed together in nearly equal amounts unusual structures referred to as ``crenellated disks'' can appear. These disks are achiral overall but the twist at the edge alternates between left- and right-handedness. To minimize the mismatch where the two regions of opposing twist meet, the ``crenellated'' structure exhibits cusps rising out of the plane of the monolayer. We use a phenomenological elastic theory previously applied to flat disks and twisted ribbons to analyze an analytic model proposed to describe the ``crenellated'' structure .  When compared with flat, circular disks, we find that the model ``crenellated disks'' are stable or at least metastable in a wide region of the phase diagram spanned by the Gaussian curvature modulus and the edge energy modulus, with a large energy barrier separating the two structures. The director pattern and geometric parameters of the ``crenellated disks'' are found to be in qualitative agreement with experimental observations.
\end{abstract}
\maketitle

\def\s{\rule{0in}{0.28in}}

\section{INTRODUCTION}
\label{intro}

The smectic \textit{A} (Sm-\textit{A}) liquid crystalline phase is a layered structure that expels twist and bend deformations analogous to the expulsion of magnetic fields in superconductors \cite{deGennes1993,deGennes1972}.  In a superconductor, magnetic fields penetrate into the bulk phase over a distance measured by the London length. Similarly, in the Sm-\textit{A} phase, two penetration depths can be defined to respectively describe the penetration of twist and bend deformations. In a Sm-\textit{A} phase composed of chiral molecules, the chirality (specifically, the chiral term in the Frank free energy) plays the role of the magnetic field in a superconductor. Chirality favors twist deformations which can appear near the edge of a smectic layer for low chirality. In the case of high chirality the twist grain boundary phase \cite{Renn1988} appears for type II smectics (i.e., those with suitably large twist penetration depth). The latter phase is the smectic analog of the Abrikosov vortex lattice in type II superconductors.

Single layer chiral Sm-\textit{A} structures have drawn significant attention recently, motivated by experimental studies of membranes formed of rodlike \textit{fd} virus particles which self-assemble in the presence of a nonbinding polymer depletant \cite{Barry2009,Gibaud2012}.  Depending on the concentration of polymer depletant the viruses self-assemble into a variety of structures including large flat disks (of order $10 \mu$m in diameter), twisted ribbons and double- and triple-helical structures formed from ribbons. In the flat disks the long axes of the viruses are aligned with the monolayer normal in the interior of the disks. Near the edge of the disk the rods tilt due to both the chiral nature of the viruses as well as the tendency to minimize the area of the virus-polymer interface. This curved edge is also observed when achiral rods are studied. For achiral viruses the spontaneous twist at the edge can be either clockwise or anticlockwise (when the membrane is viewed from above), while for chiral viruses the sense of the twist is naturally determined by the handedness of the virus. Birefringence measurements yield a twist penetration depth of approximately $0.5\mu$m \cite{Barry2009}.

\begin{figure}
\centering
\includegraphics[width = 3.0in]{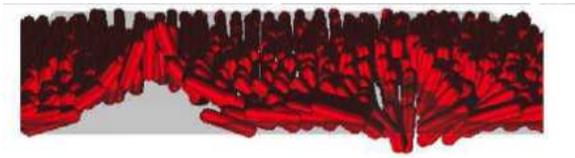}
\caption{(Color online) Schematic edge-on illustration of the arrangement of $\textit{fd}$ rods in the vicinity of a cusp at the edge of a ``crenellated disk''. (courtesy of Ref.~\cite{Gibaudprivate})}
\label{schematic}
\end{figure}

Recently, a new self-assembled \textit{fd} structure was experimentally discovered \cite{DogicDiscussion} when right-handed \textit{fd}-Y21M viruses and left-handed wild-type \textit{fd} viruses were mixed together in proportions such that the mixture is nearly achiral. Fluorescence images revealed that the two types of viruses mix uniformly without any phase separation. However, unlike the chiral case where flat membranes were observed, in the achiral limit the membranes are flat in the interior but exhibit a series of cusps along the edge where the surface of the membrane rises out of the plane of the monolayer, with adjacent cusps alternating above and below the plane. Between neighboring cusps the membrane exhibits a small radial bulge within the plane of the layer. These structures have been termed ``crenellated disks'' \cite{DogicDiscussion}.  Images produced using the 3D-LC-PolScope \cite{Oldenbourg2008} indicate that the twist at the membrane edge between adjacent cusps is of uniform chirality, alternating between right- and left-handed as the edge is encircled. The cusps are defects in the director field where right and left handed twist meet. The membrane rises out of the plane to reduce the director mismatch. Minimizing director mismatch then requires that adjacent cusps alternate above and below the plane of the disk as shown in the schematic drawing, Fig.~\ref{schematic}.

Theories describing the large flat disks and twisted ribbons have been constructed \cite{Barry2009,Pelcovits2009,Kaplan2010} using the de Gennes model for the Sm-\textit{A} phase generalized to include chirality, and in addition, in the case of the ribbons, the Helfrich model \cite{Helfrich1973,Helfrich1975} for the surface bending energy. These theories use a simple form for the edge energy, the interaction of the rods at the edge with the polymer depletant, proportional to the edge length. A more realistic model incorporating surface tension and the ``melting'' of the smectic order at the edge has recently been developed \cite{KaplanMeyer}. When applied to twisted ribbons the theory with the simple edge model yields good qualitative agreement with experimental measurements of the ribbon's pitch to width ratio providing that the Gaussian curvature modulus appearing in the Helfrich energy is positive, in contrast to the negative values typically measured in lipid monolayers or bilayers \cite{Marsh}. By comparing the free energy per unit area of a twisted ribbon with the corresponding energy of a large, flat membrane, a first-order phase transition between the two structures was found in agreement with experimental observation. However, the predicted value for the edge energy modulus was found to be an order of magnitude lower than that measured experimentally, presumably due to the very simple nature of the edge energy.

In this paper we apply the above theory to an analytic model  \cite{MeyerDiscussion} proposed to describe ``crenellated disks''. Because of the complexity of the shape we do not directly solve the Euler-Lagrange equation obtained by minimizing the free energy as was done in the case of twisted ribbons. Rather we use Monte Carlo (MC) simulations and determine the geometric parameters of the model disk and the relative stability of ``crenellated'' and flat, circular disks. We explore the stability of the ``crenellated disk'' model as a function of the Gaussian curvature and edge energy moduli. We find regions of this space where the ``crenellated disk'' is either stable or metastable. In the latter case there is a large energy barrier separating it from the flat disk. The geometric parameters for the ``crenellated disk'' throughout the phase diagram are found to be in qualitative agreement with experimental observations, as is the director pattern of the \textit{fd} viruses.

The work is organized as follows: in the next section we review the theory used earlier to study twisted ribbons and apply it to the ``crenellated disk'' model.  In Sec.~\ref{results} we present the results of our MC analysis of the free energy of the ``crenellated disk''.  We offer some concluding remarks in the final section.

\section{Free Energy of Sm-\textit{A} ``crenellated disks''}
Our analysis of the free energy of a Sm-\textit{A} ``crenellated disk'' uses the achiral limit of the theory presented in  Ref.~ \cite{Kaplan2010} for twisted ribbons. The free energy $F$ of the monolayer is the sum of the Helfrich bending energy \cite{Helfrich1973, Helfrich1975}, the de Gennes energy \cite{deGennes1972} for the Sm-\textit{A} order and director fluctuations and a simple edge energy proportional to the edge length. Specifically, we have:
\begin{equation}
F=\int(f_H+f_n)\,dA+\gamma\oint dl
\label{equ:FETotal}
\end{equation}
where $f_H$ and $f_n$ are the Helfrich and de Gennes free energy densities respectively and $\gamma$ is the edge energy modulus (bare line tension). The Helfrich free energy density is given by
\begin{equation}
f_H=\frac{1}{2}k(2H)^2+\bar{k}K_G
\label{equ:FHelfrich}
\end{equation}
where $H$ and $K_G$ are the mean and Gaussian curvature of the surface respectively, $k$ is the bending rigidity and $\bar{k}$ is the Gaussian curvature modulus. We have assumed a zero spontaneous curvature because of the up--down symmetry of the system.

The achiral de Gennes free energy density in the one-elastic constant approximation and with the assumption of perfect smectic order is given by \cite{Kaplan2010}:
\begin{equation}
f_n=\frac{1}{2}K[(\nabla\cdot\mathbf{n})^2+(\nabla\times\mathbf{n})^2]+\frac{1}{2}C\sin^2\theta \label{equ:FdGOneConstant}
\end{equation}
where $\theta$ is the relative tilt angle of the director with respect to the local surface normal, $K$ is the single Frank elastic constant and $C$ is a tilt free energy modulus. The twist penetration depth is given by $\lambda_t=\sqrt{K/C}$.

A monolayer of general shape can be modeled mathematically as a two-dimensional surface embedded in three dimensions. The surface is given by a position vector $\mathbf{Y}(u_1,u_2)$ parameterized by two coordinates, $u_1$ and $u_2$. To calculate the free energy of the membrane we use the following geometric quantities \cite{OuYang1990, OuYang1989}:
\begin{align} \mathbf{Y}_i=\partial_i&\mathbf{Y},\,\,\,\mathbf{Y}_{ij}=\partial_i\partial_j\mathbf{Y}=\Gamma_{ij}^k\mathbf{Y}_k
+L_{ij}\hat{\mathbf{N}},\nonumber\\ &\partial_i\hat{\mathbf{N}}=\partial_N\mathbf{Y}_i=-L_{ij}g^{jk}\mathbf{Y}_k,\nonumber\\
g_{ij}=\mathbf{Y}_i&\cdot\mathbf{Y}_j,\,\,\,g^{ij}=(g_{ij})^{-1},\,\,\,g=\det g_{ij},\nonumber\\
L_{ij}=\mathbf{Y}_{ij}&\cdot\hat{\mathbf{N}},\,\,\,L^{ij}=(L_{ij})^{-1},\,\,\,L=\det L_{ij}
\label{equ:DefTensors}
\end{align}
where $\partial_i\equiv \partial_{u_i}$, and $\partial_{N}$ denotes the partial derivative in  the normal direction $\hat{\mathbf N}$.  The indices $i,j,k=1,2$ and we sum over repeated indices.   The  tensors $g_{ij}$ and $L_{ij}$ are the first and second fundamental forms of the surface, respectively. The Christoffel symbols $\Gamma^k_{ij}$ are defined by the relation $\Gamma^k_{ij}=g^{km} \mathbf{Y}_{ij}\cdot\mathbf{Y}_m$.  The unit normal vector of the surface is given by
\begin{equation}
\label{eq:5}    
\hat{\mathbf N}=\frac{\mathbf{Y}_1\times\mathbf{Y}_2}{\sqrt{g}}.
\end{equation}
The Gaussian and mean curvatures are given by:
\begin{equation}
\label{eq:6}
K_G=\frac{L}{g},\,\,\,H=\frac{1}{2}g^{ij}L_{ij}
\end{equation}
and the surface area element is given by $dA=\sqrt{g} du_1 du_2$. The three--dimensional gradient operator is given by
\begin{equation}
\label{eq:12}        
\nabla=g^{ij} \mathbf{Y}_i \partial_j+\hat{\mathbf N}\partial_N\,.
\end{equation}
The director field $\mathbf{n}$ can be expressed in a local basis formed by $\mathbf{Y}_1$, $\mathbf{Y}_2$ and $\hat{\mathbf{N}}$; the director $\mathbf{n}$ makes an angle $\theta$ with $\hat{\mathbf{N}}$.

After some calculation using Eqs.~(\ref{equ:FHelfrich})-(\ref{eq:12}), the total free energy Eq.~(\ref{equ:FETotal}) is found to be:
\begin{widetext}
\begin{align}
F=&\frac{K}{2}\int
\left\{(\partial_jn_j+n_l\Gamma_{jl}^j)^2+\left(\frac{\epsilon_{3ji}}{\sqrt{g}}[(g_{ik}\partial_jn_k
+g_{il}n_k\Gamma_{jk}^l)\hat{\mathbf{N}}-(2n_kL_{jk}+\partial_j\cos\theta)\mathbf{Y}_i]\right)^2\right\}\sqrt{g}
\,du_1du_2\nonumber\\&+\frac{C}{2}\int \sin^2\theta\sqrt{g}\,du_1du_2+\bar{k}\int\frac{L}{g}\sqrt{g}\,du_1du_2
+\frac{k}{2}\int(g^{ij}L_{ij})^2\sqrt{g}\,du_1du_2+\gamma\oint\,dl,
\label{equ:FE}
\end{align}
\end{widetext}
where $\epsilon_{ijk}$ is the antisymmetric Levi--Civita tensor.

Henceforth, dimensionless units will be used where lengths are measured in units of the penetration depth $\lambda_t$ and energies are measured in units of the Frank constant $K$. These units correspond to choosing $K=C=1$ in Eq.~(\ref{equ:FE}).

We now apply Eq.~(\ref{equ:FE}) to a model \cite{MeyerDiscussion} of the ``crenellated disk'' structure observed in experiments.  We assume that the disk lies in the $x\text{--}y$ plane with an array of cusps at the monolayer edge. As noted in Sec.~\ref{intro}, minimizing director mismatch at the cusps requires that neighboring cusps have heights $h(x,y)$ (in the $z$ direction) of opposite signs. Likewise the alternation of the chirality of the twist at the edge means that cusps must appear in pairs.

We model the cusps as local peaks or valleys with an exponential decay of the height $|h|$  from the center of the cusp  \cite{MeyerDiscussion}. Specifically, for a single cusp centered at $\mathbf{r_i}=(x_i,y_i)$,  we assume that the height is given by
\begin{equation}
h_i(\mathbf{r}-\mathbf{r_i})=A_z\exp\left(-\frac{|\mathbf{r}-\mathbf{r_i}|}{b}\right)
\label{equ:Defects}
\end{equation}
where  $A_z$ is the maximum height at the peak ($A_z$ is negative for cusps lying below the plane of the disk), $b$ is a characteristic length governing the decay of the cusp into the flat central portion of the disk, and $\mathbf{r}$ is the position vector in the $x\text{--}y$ plane. For a disk with $n$ cusps ($n$ even) the height at any point $\mathbf{r}$ is given by the superposition
\begin{equation}
h(\mathbf{r})=\sum_{i=1}^n h_i(\mathbf{r}-\mathbf{r_i})
\label{equ:TotalHeight}
\end{equation}
We choose the origin of $\mathbf{r}$ at the center of the disk and the cusps are located at $\mathbf{r_i}= R (\cos{\phi_i},\sin{\phi_i})$ where $R$ is the radius of the disk (excluding the bulges) and $\phi_i$ is the polar angle location of cusp $i$.

\begin{figure}
\centering
\includegraphics[width=3.0in]{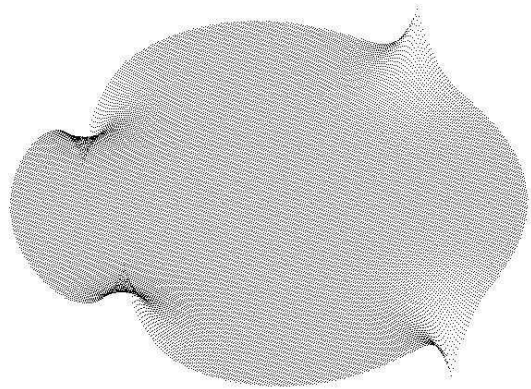}
\caption{An example of the ``crenellated disk'' model with radius $R=5.0$,$\,A_z=1.0, \,b=0.5$ (see Eq.~(\ref{equ:Defects})), and $\,A_r=0.5$ (see Eq.~(\ref{equ:Edges})) in units of the penetration depth. The four cusps are distributed uniformly along the perimeter of the disk. Note that the director field is not displayed in this figure.}
\label{fig:Shape}
\end{figure}

We model the in-plane radial bulges between neighboring cusps labeled by $i$ and $i+1$ as follows
\begin{equation}
R_i(\phi)=R+A_r\sin\left(\frac{\pi(\phi-\phi_i)}{\phi_{i+1}-\phi_i}\right)
\label{equ:Edges}
\end{equation}
where $R_i$ specifies the radial coordinate of the disk edge between the cusps and $A_r$ is the magnitude of the protrusion. The edge of the disk is then given as a piecewise function of these protrusions between neighboring cusps.
Fig.~\ref{fig:Shape} shows an example of the shape of the ``crenellated disk'' model of Ref.~\cite{MeyerDiscussion}.

With an analytic form of the shape specified, the differential geometry quantities defined in Eq.~(\ref{equ:DefTensors}) can be computed explicitly as functions of $x$ and $y$ and substituted into the free energy Eq.~(\ref{equ:FE}). In principle we could then follow the approach of Ref.~\cite{Kaplan2010} and derive the corresponding Euler-Lagrange equations for the director field. However because of the complexity of the shape, the resulting equations are impossible to solve explicitly even using numerical solvers. Instead we discretize the underlying $x\text{--}y$ plane using a square lattice of grid size $0.05$ and carry out an MC simulation at low temperature ($10^{-4}$ in dimensionless energy units with $k_B=1$) varying the geometrical parameters of the shape. We note that our dimensionless energy unit, $K=1$, corresponds to approximately $100 k_B T$ at room temperature using the measured value of the twist elastic constant in $\textit{fd}$ solutions \cite{Dogic1}. Thus, thermal fluctuations of the director are negligible, as noted already in Refs.~\cite{Barry2009, Kaplan2010}, where the free energy was minimized and very good agreement was found between the predicted director pattern and experimental measurements. We initialized our system with both a random and fully aligned director field and found similar equilibrium states. We also carried out a multi-grid computation where the grid is finer at the edge where the directors are twisted and similar results were obtained.

Experimentally \cite{Gibaud2012} the director field at the edge of the disk is observed to be tangent to the edge. We imposed this boundary condition in our model by introducing ghost directors along the edge of the ``crenellated disk''. These directors are fixed tangent to the edge and interact with neighboring directors in the interior of the disk via the Frank free energy. We used a central difference algorithm to compute the derivatives of the director field. The total free energy was computed using numerical integration over all the lattice sites inside the  disk. Director defects can appear at the cusps and we include a defect core energy $\sum K_{def}\theta_z^2$ where $\theta_z$ is the angle made by the director with respect to the z-axis. The summation runs over all the lattice sites whose distances from a cusp in both the $x-$ and $y$-directions are less than one lattice constant. The energy $K_{def}$ is chosen to be equal to the Frank constant $K$ which is unity in our system of units.

Experimentally ``crenellated disks'' are observed to maintain constant area after their formation and thus as we varied the geometric parameters of the shape we adjusted the radius $R$ so that the total area of the disk is kept constant. For each set of geometric parameters we carried out $100000$ MC cycles where each director is allowed to move once during each cycle and data are collected over the last $50000$ cycles. The step size of the test move was chosen so that the overall acceptance ratio is approximately $50\%$.

Since we are effectively solving the Euler-Lagrange equations whose solution corresponds to the zero-temperature ground states of the system, we collected the orientations of each director in each cycle and computed the averaged orientations for each of them and then used this averaged configuration to compute the free energy instead of the usual method that collects the free energy every cycle and then averages that quantity. We have checked the validity of this method by considering a flat circular disk and obtaining good agreement between the results of our MC simulations and the results obtained numerically in Ref.~\cite{Pelcovits2009} where the Euler-Lagrange equation was solved explicitly.

\section{RESULTS}
\label{results}
We consider a ``crenellated disk'' with  four uniformly distributed cusps. The area of the disk is approximately 100 corresponding to a radius of approximately 5 in our dimensionless units. Significantly larger disks require computational resources beyond what are available to us. The decay length $b$ is chosen to be 0.2, of the order of magnitude observed in experiments.
\begin{figure}
\centering
\includegraphics[width=2.5in]{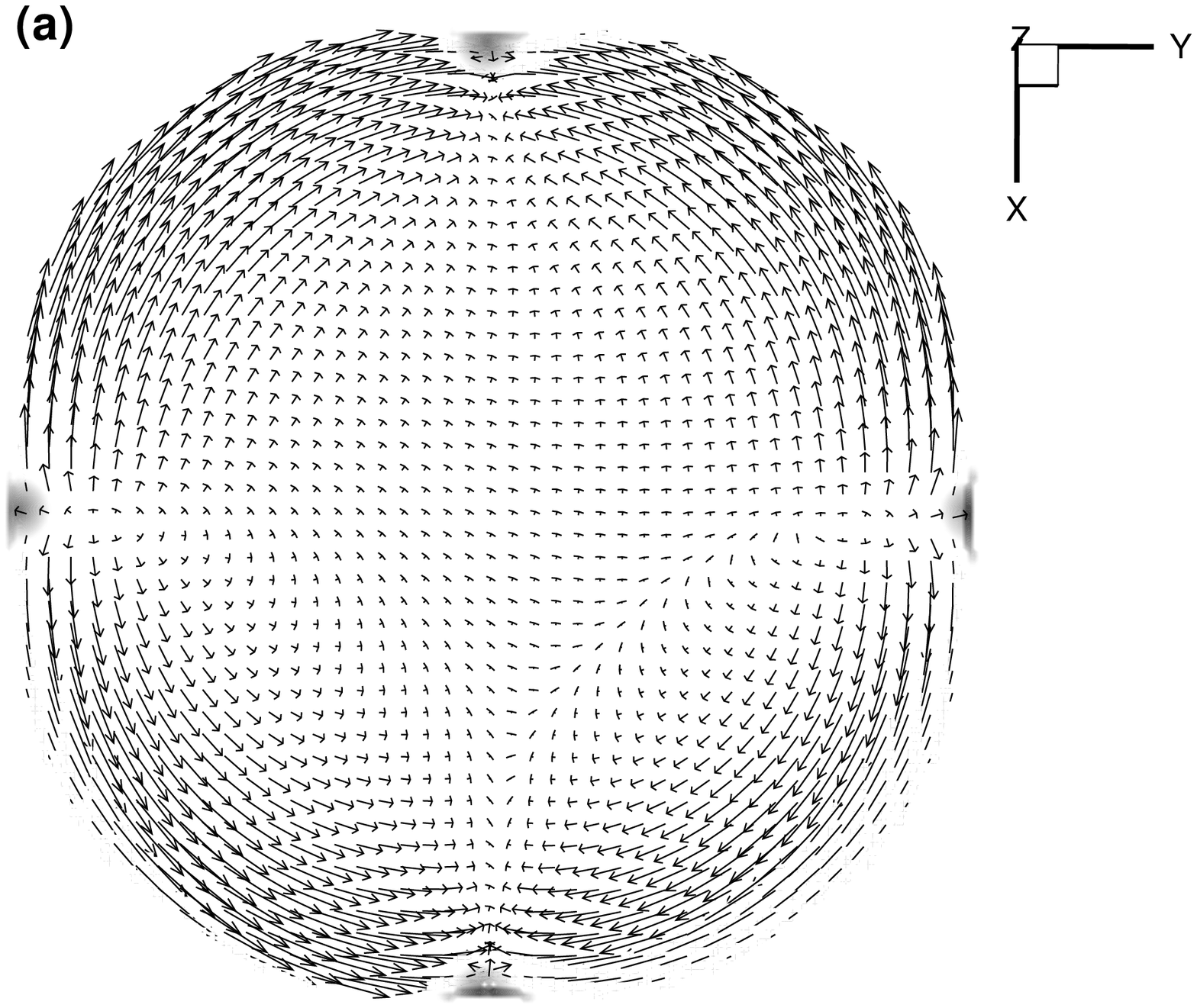}
\includegraphics[width=2.5in,angle=270]{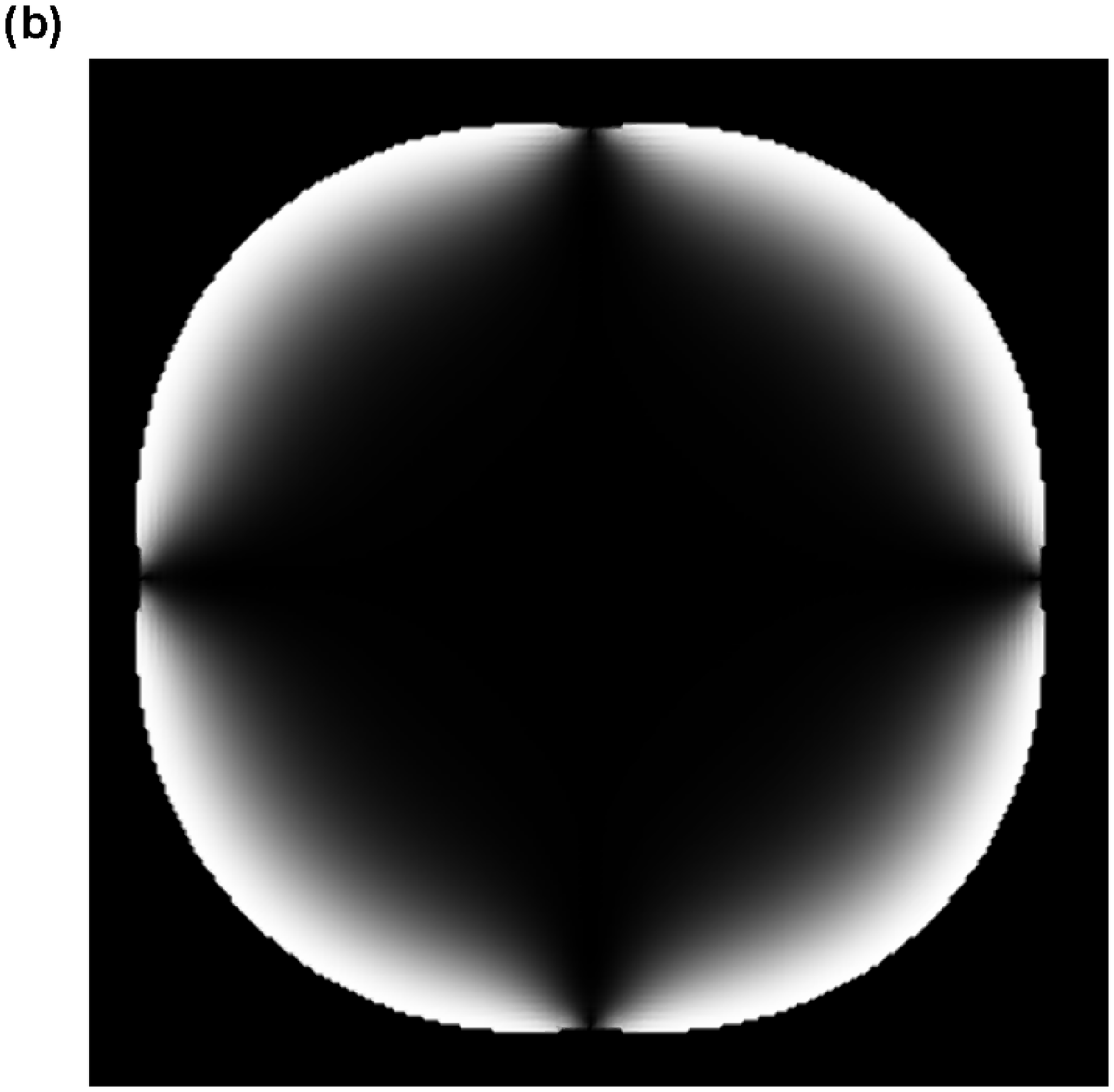}
\caption{(a)An example of the simulated director field on a ``crenellated disk'' with $R=5.41,\,A_z=0.6,\,A_r=0.4,\,b=0.2$ in units of the penetration depth with four cusps distributed uniformly along the edge. (b)The corresponding simulated birefringence image where the gray scale is proportional to $\sin^2\theta$; $\theta$ is the angle the director makes with the $z$ axis.}
\label{Fig:Solution}
\end{figure}
An example of the simulated director field on a ``crenellated'' structure is shown in Fig.~\ref{Fig:Solution}(a). Near the center of the disk the directors are approximately perpendicular to the $x\text{--}y$ plane as expected because of the tilt energy term in the free energy. The boundary condition at the edge forces the directors there to lie tangent to the edge. It is seen from the figure that there is a defect in the director field at each cusp where the chirality of the director tilt near the edge changes sign from one side of the cusp to the other. The director twist on either side of the cusp relaxes over a length scale of order the penetration depth.  Fig.~\ref{Fig:Solution}(b) shows the simulated birefringence image where the gray scale is proportional to $\sin^2\theta$. The  bright regions correspond to the radial bulges of the membrane where the directors tilt more while the dark regions correspond to the cusps where the director is nearly vertical (i.e., pointing in the $z$-direction). This is qualitatively in agreement with experimental observations \cite{DogicDiscussion}.

\begin{figure}
\centering
\includegraphics[width=2.0in]{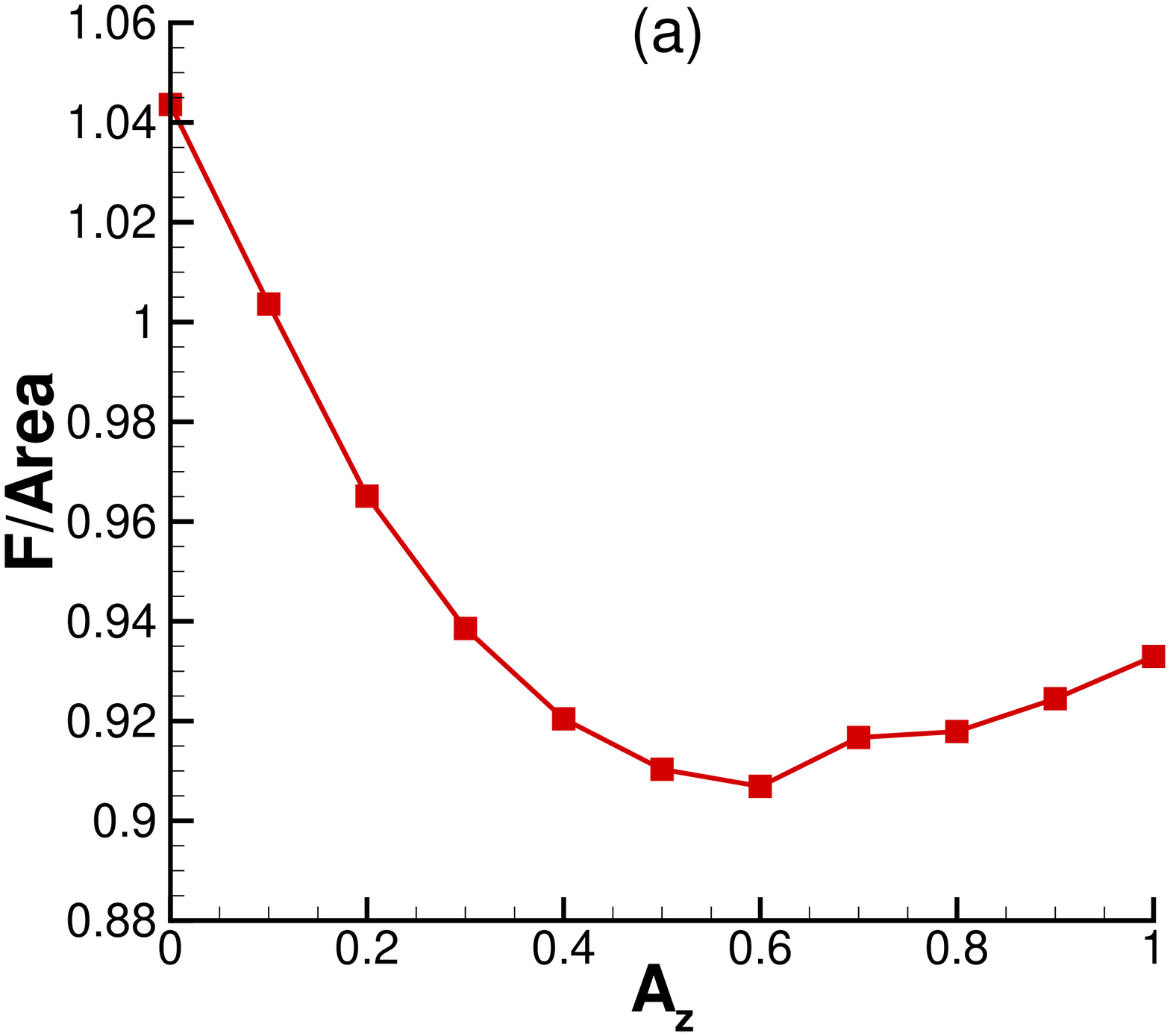}
\includegraphics[width=2.0in]{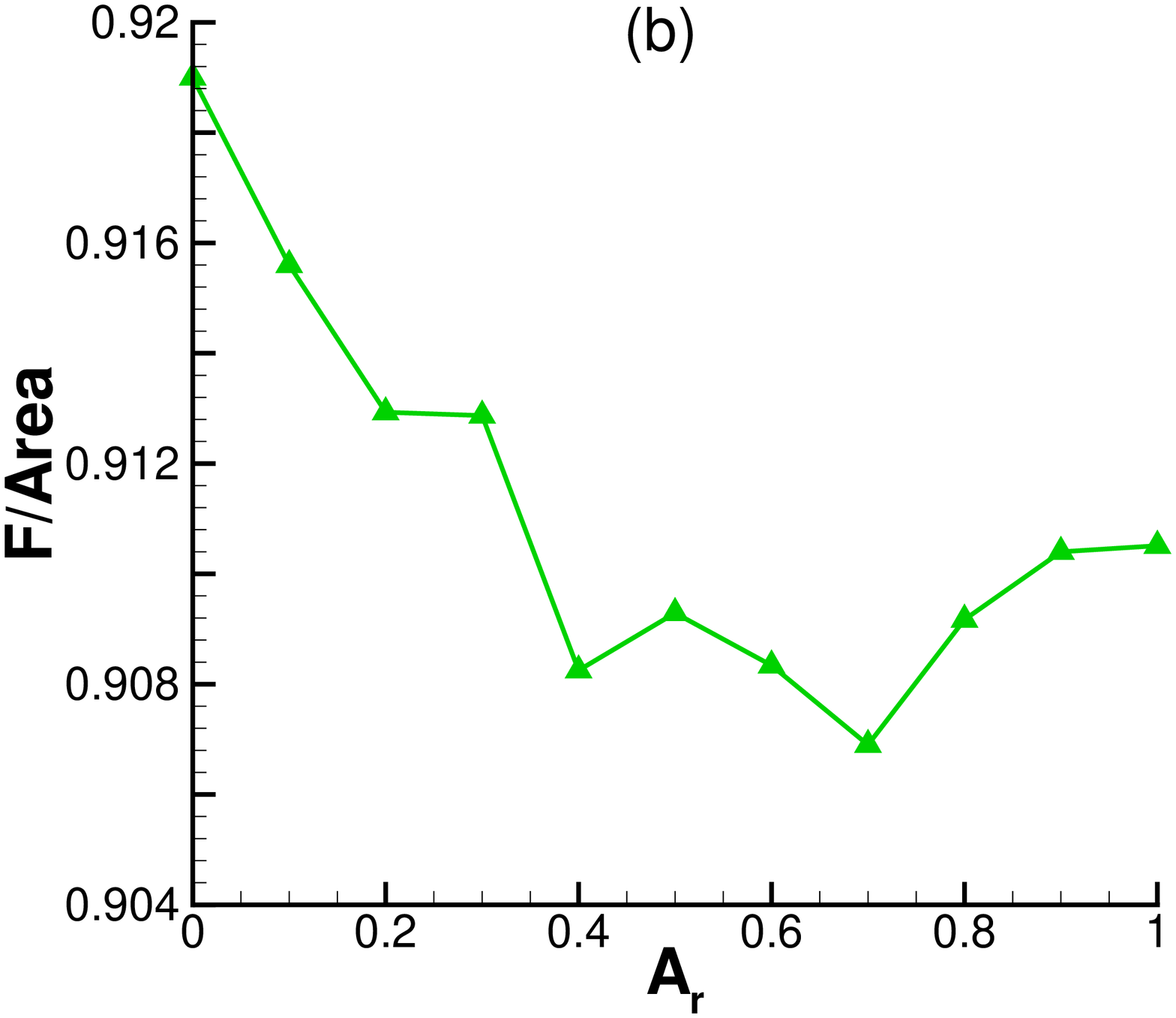}
\includegraphics[width=2.0in]{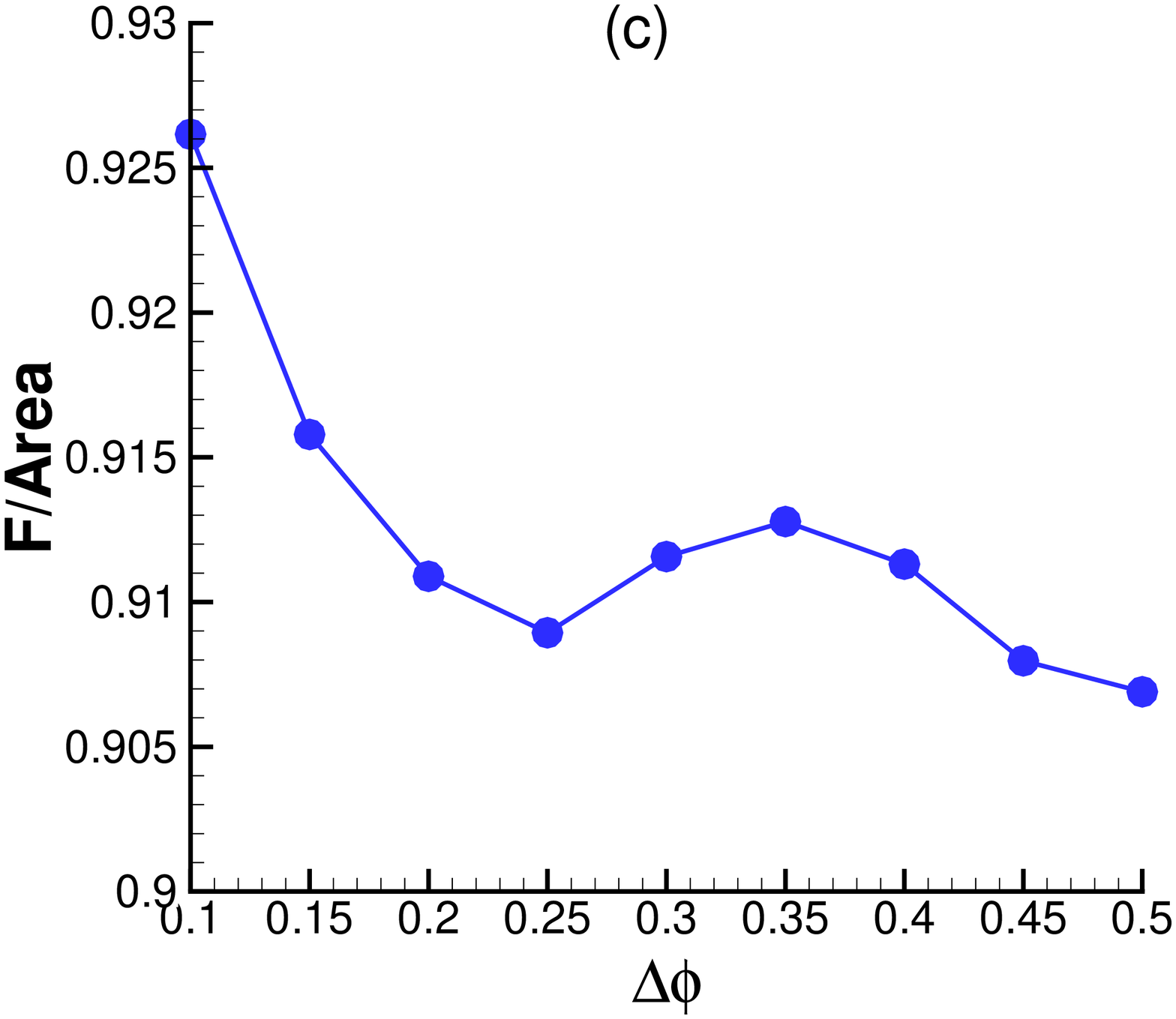}
\caption{(color online)(a) Free energy per unit area of a ``crenellated disk'' as a function of height of the cusps $A_z$ with  $A_r=0.7$, $b=0.2$, $\bar{k}=1.5$, $\gamma=1.6$, and a uniform distribution of the cusps along the edge of the disk. (b)  Free energy per unit area of the disk as a function of the size of the radial bulges $A_r$ with  $A_z=0.6$, and $\bar{k}$, $b$ and $\gamma$ fixed as in (a). (c) Free energy of the disk as a function of the angular separation between adjacent cusps with  $A_z=0.6$, $A_r=0.7$, and $\bar{k}$, $b$ and $\gamma$ fixed as in (a). The angular separation is given in units of $\pi$; e.g., a separation of 0.4 means that the four cusps are placed at $\phi=0.0, 0.4\pi, 0.8\pi, 1.2\pi$. Uniform distribution of the cusps corresponds to $\Delta\phi=0.5$.  The lines in all of the figures are guides to the eye.}
\label{Fig:Behcookie}
\end{figure}

\begin{figure*}
\centering
\includegraphics[width=2.0in]{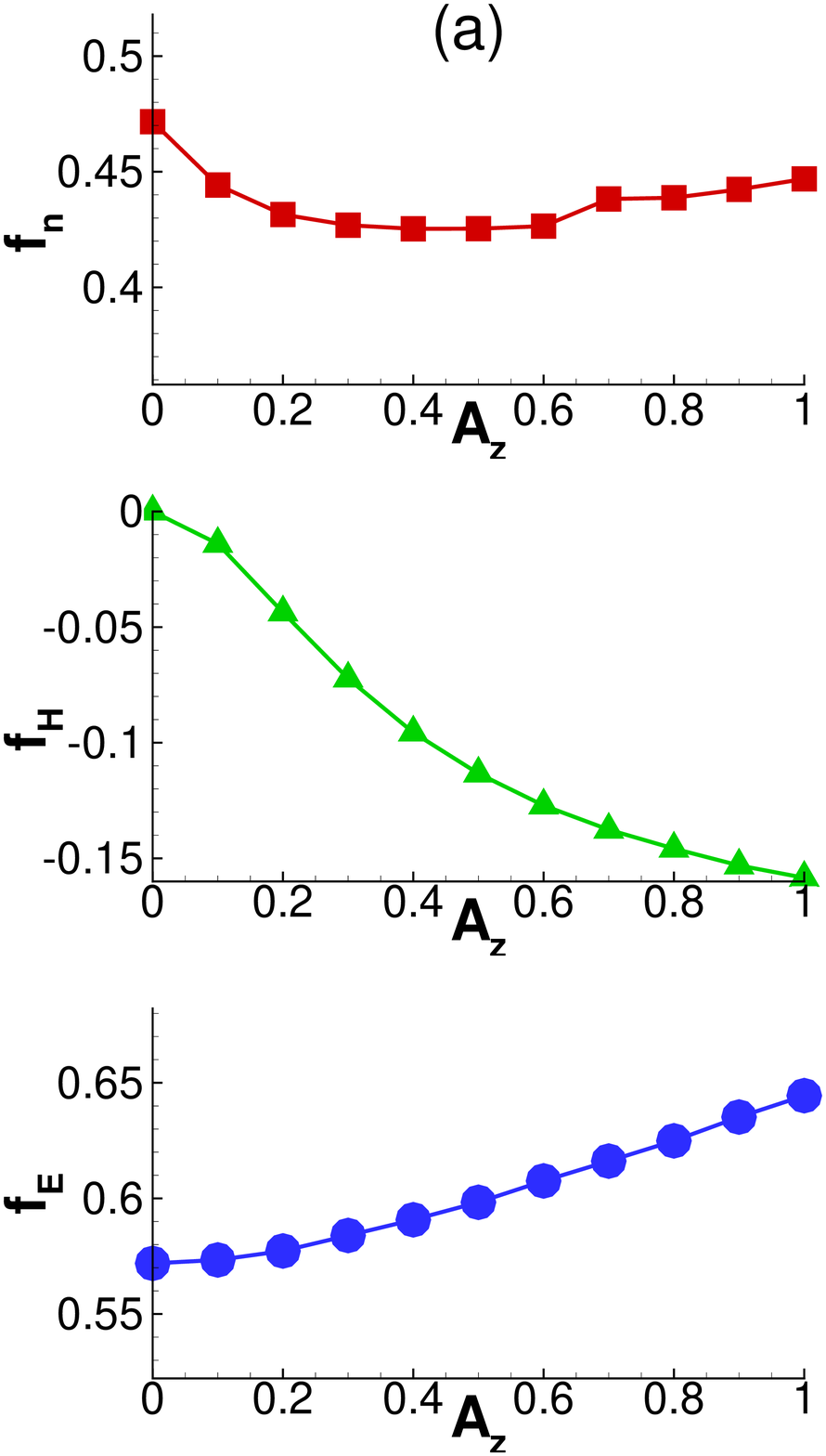}
\includegraphics[width=2.0in]{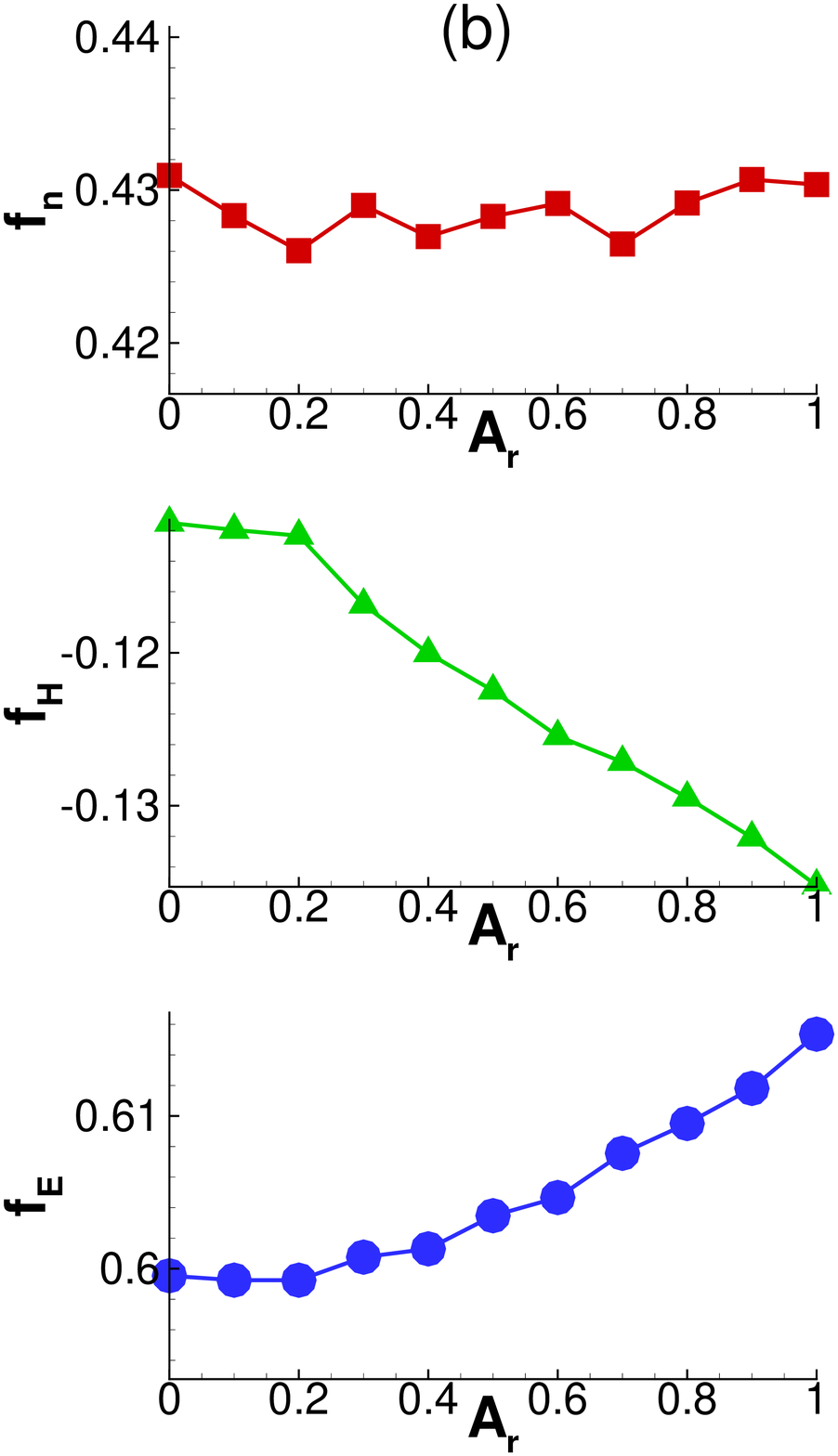}
\includegraphics[width=2.0in]{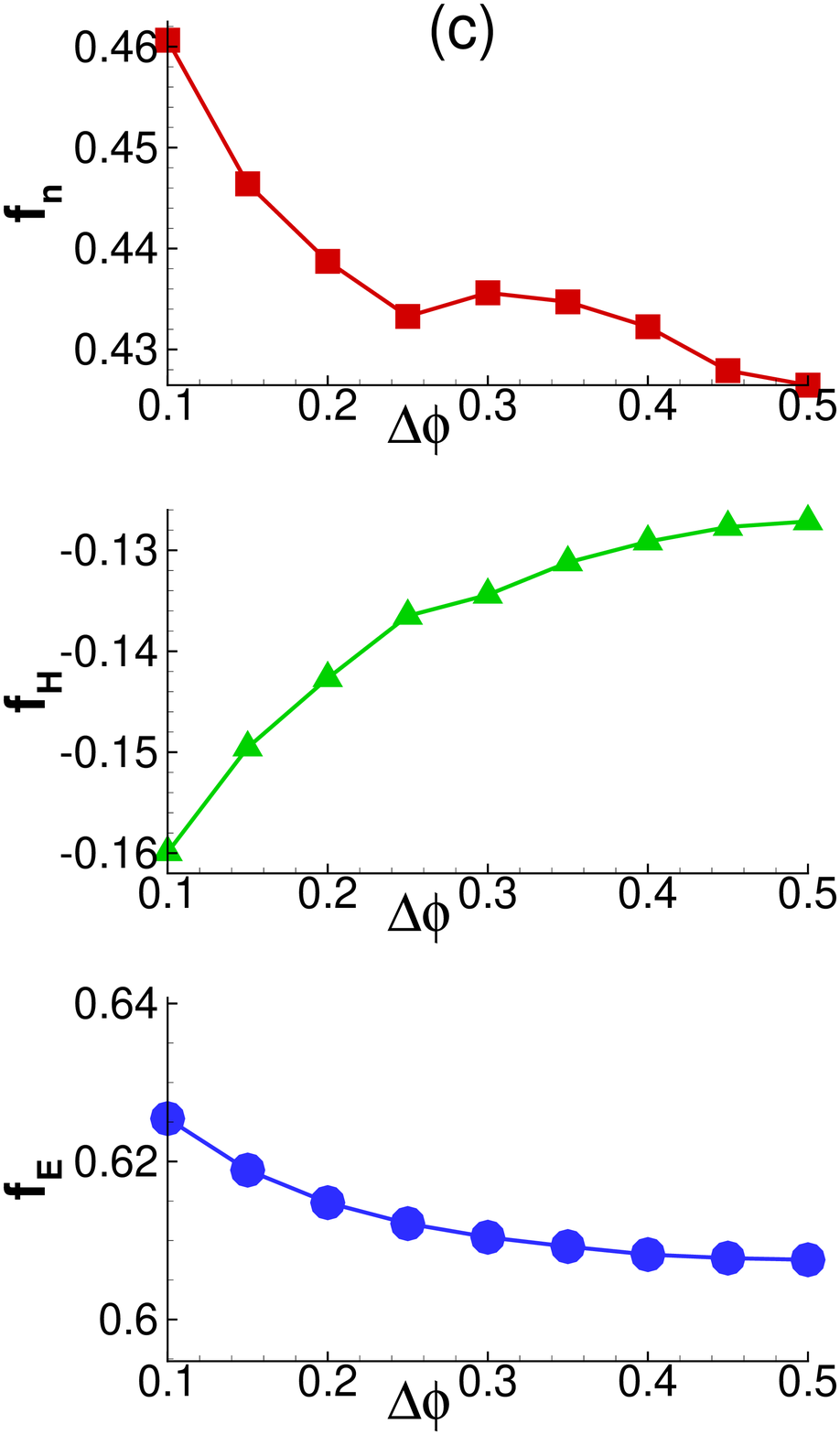}
\caption{(color online)Contributions of the Frank ($f_n$), Gaussian curvature ($f_H$) and edge ($f_E$) energy densities to the total free energy per unit area for the ``crenellated disk'' as functions of (a) the height of the cusps $A_z$, (b) the size of the radial bulges $A_r$, and (c) the angular spacing $\Delta \phi$ (measured in units of $\pi$) between adjacent cusps. As in Fig.~\ref{Fig:Behcookie} we have set $\bar{k}=1.5$ and $\gamma=1.6$. Note that the ranges of the y-axes in the three figure parts are different but the spans of energy are equal, allowing easy comparison of the magnitudes of the variation of each term. }
\label{Fig:Termcookie}
\end{figure*}

To understand why these achiral membranes would prefer to form cusps and bulges instead of remaining flat and circular, we analyze the contributions from different terms to the total free energy when $A_z$ or $A_r$ is varied. In Fig.~\ref{Fig:Behcookie}(a), the variation of the total free energy per unit area with the height of the cusps $A_z$ is shown with $A_r=0.7$ and the cusps are distributed uniformly along the edge. We have chosen $\bar{k}=1.5$ and $\gamma=1.6$; our results, however, are representative of a wide range of values for these parameters as we discuss below. We have assumed that the mean curvature modulus $k$ is zero to simplify our analysis. Neither $k$ nor $\bar{k}$ has been measured experimentally in the $\textit{fd}$ systems but experimental evidence suggests that it is the Gaussian curvature that plays a dominant role. In the case of twisted ribbons \cite{Kaplan2010} a small second-order Gaussian curvature term was added to the Helfrich energy in order to stabilize the energy. In the present case the negative Gaussian energy proportional to $\bar{k}$ does not lead to any instabilities and thus we have not included a higher order term in our analysis.  Contributions from the different terms in the free energy are shown in Fig.~\ref{Fig:Termcookie}(a).  It is seen that the Frank free energy assumes a parabolic shape like the total free energy, i.e., there is a minimum at a nonzero value of $A_z$, because the out-of-plane cusp structure relaxes the mismatch of the directors in the right-handed and left-handed regions flanking the cusp. This optimal value of $A_z$ is not arbitrarily large because a large value would lead to a large deformation in the director field in the neighborhood of the cusp. It is also seen that both the absolute value of Gaussian curvature (note that Gaussian curvature is always negative in this structure) and the length of the edge increase monotonically with $A_z$. For low values of $A_z$, the Frank energy dominates and at large $A_z$, both the edge and Frank free energy combine to overcome the gain in Gaussian curvature free energy. Thus, a minimum in the total free energy is produced at  $A_z\approx 0.6$ for the the given values of $\bar{k}$ and $\gamma$. .

The behavior of the free energy when the radial extent, $A_r$, of the in-plane radial bulge between neighboring cusps is varied is shown in Fig.~\ref{Fig:Behcookie}(b) with $A_z=0.6$, the value found to minimize the free energy. The parameters $\bar{k}$ and $\gamma$ were again chosen to be 1.5 and 1.6 respectively. Compared to the dependence of the free energy on the out-of-plane height, $A_z$, of the cusps, the dependence on the size of the radial bulges is much weaker, possibly because the cusps are sharply defected structures while the radial bulges are much smoother. The contributions to the free energy from the Frank, Gaussian curvature and edge energies are shown in Fig.~\ref{Fig:Termcookie}(b). Larger bulges introduce larger deformations in the director field far from the cusps because the directors are forced to follow a more curved edge at the boundary, but larger bulges can also reduce the mismatch between the directors near the cusps by forcing the directors on both sides of the cusps to become more parallel. These two competing effects appear to essentially cancel each other, and the Frank free energy is only weakly dependent on $A_r$, with some numerical fluctuations. The absolute value of the Gaussian curvature increases with $A_r$. The edge energy initially decreases slightly with increasing $A_r$ and then for $A_r\gtrsim 0.15$, the edge energy increases. The tradeoff between increasing edge energy and the gain of Gaussian curvature energy leads to a free energy minimum at $A_r\approx  0.7$.

In experiments on large ``crenellated disks'' cusps have been observed to lie on only a portion of the disk's edge rather than being distributed along the entire edge of the membrane. We consider this possibility for our small disks as shown in Fig.~\ref{Fig:Behcookie}(c) where we plot the free energy per unit area as a function of the angular separation between the cusps, assuming $A_z=0.6$, $A_r=0.7$, $b=0.2$, $\bar{k}=1.5$ and $\gamma=1.6$. The separation is plotted in units of $\pi$ and thus a separation of 0.5 corresponds to the configuration where the four cusps are placed uniformly along the edge. The contributions to the total energy per unit area from the Frank, Gaussian curvature and edge energies are shown in Fig.~\ref{Fig:Termcookie}(c). While squeezing the cusps into a small segment of the edge creates a flatter segment elsewhere which reduces the Frank free energy there, it also creates a more curved geometry in the region where the cusps are located which increases the Frank free energy in that location. From the general trend of the Frank free energy shown in the figure, we see that the latter effect overcomes the former so that the Frank free energy increases as the cusps are brought together. As expected, the edge length and the Gaussian curvature both increase in  absolute value upon squeezing the cusps together because the membrane is more curved in the squeezed region and this region provides the main contribution to these two terms compared to the flatter region. For the $\bar{k}$ and $\gamma$ values selected, the trend of the change in the free energy is dominated by the Frank free energy and is minimized when the four cusps are placed uniformly as shown in Fig.~\ref{Fig:Behcookie}(c). We are unable to explain the appearance of another minimum at spacing 0.25 which may be due to the numerical fluctuations in the Frank free energy term. It appears that our disks are simply too small to exhibit cusps lying on a portion of the disk perimeter as seen in experiments.

Experimental observations \cite{DogicDiscussion} indicate that ``crenellated disks''  with $A_z$ comparable to but smaller than the twist penetration depth should be at the very least metastable structures and possibly true equilibrium shapes under certain physical conditions. We now address the stability and metastability of the theoretical ``crenellated disk'' model. We consider the energetics of the disk as both the Gaussian curvature modulus $\bar{k}$ and the edge energy modulus $\gamma$ are varied.
\begin{figure}
\centering
\includegraphics[width=3.0in]{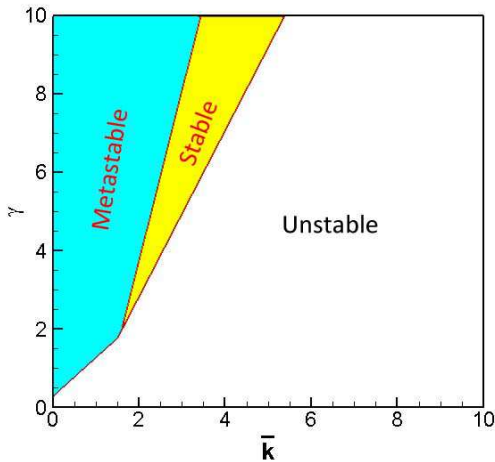}
\caption{(color online)The phase diagram for the ``crenellated disk'' in the $\bar{k}\textit{--}\gamma$ space. ``Metastable'' and ``stable'' in this figure refer to the ``crenellated disk'' in comparison with a perfectly flat, circular disk of the same area. In the unstable region either $A_z$ or $A_r$ becomes very large, indicating an instability toward other geometrical structures beyond the scope of our model.}
\label{Fig:PD}
\end{figure}

Our results are shown in Fig.~\ref{Fig:PD}. ``Metastable'' and ``stable'' in this figure refer to the ``crenellated disk'' in comparison with a perfectly flat, circular disk of the same area. Throughout the metastable and stable regions shown, $A_z$ is approximately one half of the twist penetration depth in accord with experiments. Our theoretical prediction for $A_r$ is also approximately one half of the twist penetration depth while experimentally larger bulges have been observed. We attribute this to the fact that our theoretical calculations have used disks smaller than the real disks observed in experiments. It is reasonable to assume that value of $A_r$ will increase with the radius $R$ of the disk. Thus, we argue that our theoretical model has produced reasonable values for both $A_z$ and $A_r$ compared to experiments. In the ``unstable" region we have found that either $A_z$ or $A_r$ becomes very large, indicating an instability toward other geometrical structures which we cannot determine in our model. To get a sense of the energetics, consider, e.g., $\bar{k}=3$. For large $\gamma$ ``crenellated disks'' are at best only metastable because the edge energy is too large to favor such a structure with a high edge-to-area ratio compared to flat disks. When $\gamma$ is lowered, the ``crenellated disk'' can become stable. However, when $\gamma$ is lowered further the ``crenellated disk'' becomes unstable to other structures with even larger edge-to-area ratios such as helices. For fixed $\gamma$ if the Gaussian curvature modulus is too large structures with greater negative Gaussian curvature can again appear in place of ``crenellated disks''.  In most of the metastable region, the energy barrier between the perfectly flat, round disks and the ``crenellated disk'' is about several Frank constants corresponding to several hundred $k_B T$ in physical units.

In Ref.~\cite{Kaplan2010} the transition from large, flat disks to twisted ribbons in the elastic theory was determined to occur at $\gamma\approx 0.28$ and $\bar{k}\approx 0.1$, corresponding to the very bottom edge of the metastable region of our phase diagram. This value of $\gamma$ is one order of magnitude lower than the experimentally measured value. The Gaussian curvature modulus has not been measured experimentally, though there is some evidence \cite{DogicDiscussion} that suggests its value is larger, possibly by one order of magnitude, than the value found in the theory of Ref.~\cite{Kaplan2010}. Thus, in our phase diagram we have explored values of $\gamma$ and $\bar{k}$ larger than those used to analyze the twisted ribbon. In the latter analysis free boundary conditions were used for the orientation of the \textit{fd} viruses at the edge of the monolayer which leads to a tilt angle $\theta$ at the edge which is significantly smaller than the $90^\circ$ measured experimentally.  In the present case we imposed a $90^\circ$ tilt at the edge both to fit the experimental observations and to incorporate a nonzero tilt in an achiral model which otherwise would have zero tilt (at least for the simple edge model used here and in Ref.~\cite{Kaplan2010}). Thus, we believe that the range of values for $\gamma$ and $\bar{k}$ shown in Fig.~\ref{Fig:PD} are reasonable to explore in comparing theory with experiment.

Given the experimental measurements to date, we have restricted our analysis to the possibility of ''crenellated disks'' with cusp heights $A_z$, and radial bulges, $A_r$, no larger than one penetration depth. We cannot exclude the possibility that ``crenellated disks'' with larger values of these parameters can be stable or metastable outside of the stable and metastable regions shown in the figure, but we have not explored this possibility due to the limitation of our computational resources. The ``crenellated'' region obtained by our current computations is quite possibly an underestimate.

Kaplan and Meyer \cite{KaplanMeyer} have recently considered a model of an array of cusps on the edge of a flat membrane in the absence of the director field. Each cusp is modeled as a generic surface of revolution with negative Gaussian curvature (this model does not include the decaying exponential of Eq.~(\ref{equ:Defects}) so as to maximize the saddle characteristics of the geometry).  By simultaneously optimizing the shape of the surface and its free edges Kaplan and Meyer obtain an analytic relation between $k$, $\bar{k}$, $\gamma$ and $A_r$, which for $k=0$ yields: $\bar{k}=\gamma A_r$. In the achiral limit their analysis of the energy of an array of cusps yields a transition to a ``crenellated disk'' with $A_r \sim 0.5$ for $\bar{k}\sim 1$ and $\gamma\sim2$, in dimensionless units, of the same order of magnitude as our results.

\section{CONCLUSION}

We have used an elastic theory of Sm-\textit{A} monolayers to study a model \cite{MeyerDiscussion} of the ``crenellated disks'' observed in achiral mixtures of \textit{fd} viruses. The theory is the achiral limit of one used earlier \cite{Barry2009, Pelcovits2009, Kaplan2010} to describe flat disks and twisted ribbons in chiral \textit{fd} monolayers. Using MC simulations we computed the geometric parameters of the ``crenellated disks'', the director field of the viruses and the relative stability of the disks. The height of the cusps was found to be in good agreement with experimental observation. Although we obtained a smaller radial protrusion compared to experiments, our theoretical prediction is still reasonable noting the smaller size of the disks we have used in our calculations. The director field is also in good agreement with experimental birefringence measurements. The ``crenellated disks'' are found to be stable in a region where the Gaussian curvature modulus $\bar{k}$ and $\gamma$ are both one order of magnitude higher than the value previously obtained when the elastic theory was applied to twisted ribbons. However, the metastable region does include the values of the moduli used to describe ribbons.

\begin{acknowledgments}

We thank E. Barry, Z. Dogic, T. Gibaud, C. N. Kaplan, R. B. Meyer, P. Sharma and M. Zakhary for helpful discussions. We are grateful to R.  B. Meyer for suggesting the analytic form of the ``crenellated disk'' used here. This work was supported by the NSF through MRSEC Grant No. 0820492.

\end{acknowledgments}

\end{document}